\documentclass[conference]{IEEEtran}

\usepackage{amsmath,amssymb}
\usepackage{graphicx}
\usepackage{cite}
\usepackage{xcolor}
\usepackage{placeins}

\usepackage{algpseudocode}
\usepackage[ruled,vlined,linesnumbered,noend]{algorithm2e}

\SetAlgoNlRelativeSize{-1}      
\DontPrintSemicolon             
\SetAlCapSkip{0.4em}            
\IncMargin{-0.5em}              
\usepackage{tikz}
\usepackage{pgfplots}
\pgfplotsset{compat=1.7}

\usepackage{etoolbox}      
\makeatletter
\patchcmd{\subsection}{\addvspace{1ex}}{}{}{}   
\makeatother

\SetKwInput{KwNote}{\textbf{Note}}
\begin{document}
%
\title{Mobility-Aware Localization in mmWave Channel: Adaptive Hybrid Filtering Approach 
\thanks{Corresponding author: orimogunjea@run.edu.ng}}
\author{\IEEEauthorblockN{Abidemi Orimogunje\IEEEauthorrefmark{1}\IEEEauthorrefmark{2}, Kyeong-Ju Cha\IEEEauthorrefmark{2}, Hyunwoo Park\IEEEauthorrefmark{2}, 
Abdulahi A. Badrudeen\IEEEauthorrefmark{2},
Sunwoo Kim\IEEEauthorrefmark{2},\\  Dejan Vukobratovic\IEEEauthorrefmark{3} 
\vspace{1mm}
\IEEEauthorblockA{\IEEEauthorrefmark{1}{African Center of Excellence in IoT,
University of Rwanda, Rwanda}
\IEEEauthorblockA{\IEEEauthorrefmark{2}Department of Electronic Engineering, Hanyang University, South Korea}
\IEEEauthorblockA{\IEEEauthorrefmark{3}Faculty of Technical Sciences, University of Novi Sad, Serbia}
}}}

\maketitle

\maketitle
\begin{abstract} 
Precise user localization and tracking enhance energy-efficient and ultra-reliable low-latency applications in the next-generation wireless networks. In addition to computational complexity and data association challenges with Kalman-filter-localization techniques, estimation errors tend to grow as the user’s trajectory speed increases. By exploiting mmWave signals for joint sensing and communication, our approach dispenses with additional sensors adopted in most techniques while retaining high-resolution spatial cues. We present a hybrid mobility-aware adaptive framework that selects between the Extended Kalman Filter at pedestrian speed and the Unscented Kalman Filter at vehicular speeds. The scheme mitigates data-association problem and estimation errors through adaptive noise scaling, chi-square ($\chi^{2}$) gating, and Rauch–Tung–Striebel smoothing. Evaluations using Absolute Trajectory Error, Relative Pose Error, Normalized Estimated Error Squared, and Root Mean Square Error metrics demonstrate roughly 30–60\,\% improvement in their respective regimes, indicating  a clear advantage over existing approaches tailored to either indoor or static settings.

\end{abstract}



\section{Introduction}
Precise localization and tracking of user equipment (UE) are key enablers for seamless mobility management and ultra-reliable communication in future 6G networks \cite{saleh2025integrated, chen2022tutorial,palacios2019single}. Leveraging the built-in sensing capability of millimeter-wave (mmWave) signals removes the need for extra sensors and thus lowers the cost for integrated sensing-and-communication (ISAC) services \cite{ge2023integrated,amjad2023radio, wan2018robust}. While many existing mmWave localization schemes focus on static or indoor settings, only a handful address highly dynamic outdoor scenarios with rapidly varying user mobility \cite{palacios2019single, shi2019anchor}. The mmWave ISAC system can operate in either monostatic or bistatic mode, depending on its antenna layout. A monostatic configuration uses the same antenna for transmission and reception, whereas a bistatic setup employs separate antennas for the transmit and receive paths \cite{ge2023integrated}.

In the bistatic mmWave radio-channel framework, a gNB transmits probing signals
whose sensing modalities; time of arrival (ToA), angle of arrival/departure (AoA/AoD), Doppler spread, and wheel/Inertial Measurement Unit (IMU) odometry are
exploited for localization and tracking. When an environment is already mapped (i.e., known landmarks and virtual anchors are available), these
features can directly aid pose estimation \cite{amjad2023radio, shi2019anchor, zheng2023simultaneous}. For unmapped areas, simultaneous localization and mapping (SLAM) may be required for accurate estimation of the user's trajectory, but for a mapped environment, only the localization technique is sufficient for estimating the position and orientation of the user. In both cases, the Bayesian filtering or batch optimization can recover both the UE pose and the
environmental layout \cite{amjad2023radio, wan2018robust,zheng2023simultaneous}.

Meanwhile, the deployment of Bayesian Kalman filters for localization in high mobility outdoor mmWave channels is
challenging as multipath and clutter hinder data association leading to mis-match of landmarks and error in estimation \cite{chen2022tutorial,julier1997new}.  Noise statistics vary with speed and propagation, so fixed process and
measurement noise settings \((Q,R)\) become inconsistent and inflate
performance metrics such as Normalized Estimated Error Square (NEES). The Extended Kalman Filter (EKF) linearizes poorly at vehicle speeds, whereas the Unscented Kalman Filter (UKF) handles the
non-linearity well but is heavier and fragile at low speeds \cite{julier1997new, kumar2023survey}.
Reflections and NLoS links further demand strict chi-square ($\chi^{2}$)\ gating to
block spurious AoA/AoD/ToA hits; otherwise, Absolute Trajectory Error (ATE) spikes occur \cite{julier2004unscented}.
These issues call for a mobility-aware scheme that can switch between appropriate filters and minimize noise in real time.

We design a \emph{hybrid, mobility-aware} localization framework with EKF and UKF using an odometry-derived speed gate. When the user's speed is $\bigl(\lVert\hat v_k \rVert\le 2\,\mathrm{m/s}\bigr)$, the system enters a low-mobility mode running the EKF model; for the speed range
$\bigl(2 < \lVert \hat v_k\rVert \le 20\,\mathrm{m/s}\bigr)$ the pipeline activates the UKF model better suited to nonlinear, high-speed motion. In both modes we employ adaptive $Q/R$ scaling, $\chi^{2}$ Mahalanobis gating,
virtual-anchor augmentation, and a Rauch–Tung–Striebel (RTS) smoother to
overcome the intrinsic Bayesian filter limitations such as the data-association and noise-mismatch issues that plague outdoor mmWave localization \cite{chen2022tutorial, julier1997new, julier2004unscented}.

Using realistic DeepMIMO ray-tracing datasets, we evaluated the performance of the localization techniques through ATE, Relative Pose Error (RPE), NEES, and Root-Mean-Square Error (RMSE) metrics\cite{scheideman2020flexible, yi2019metrics}. 
By coupling adaptive noise control, statistical gating, virtual reflectors, and RTS smoothing, the hybrid model maintains sub-$0.25\,\text{m}$ ATE for pedestrians and
$\approx 2\,\text{m}$ ATE for vehicles showing significance improvement compared to the state-of-the-art solutions. Particularly, the indoor mmWave localization methods in
\cite{karttunen2025towards} report $0.2$–$0.3\,\text{m}$ ATE and the RadarSLAM in adverse outdoor conditions suffers $5$–$13\,\text{m}$ ATE drift\cite{hong2020radarslam}. This indicates an improvement over the existing model, and our approach provides a practical path toward mobility-aware beam management in next-generation networks\cite{shahmansoori2017position}.


In summary, the core contributions of this paper include:
\begin{enumerate}
\item Optimization of Bayesian Kalman filters through adaptive noise scaling, chi-square gating, and RTS smoothing techniques for accurate track estimation
\item Development of filter-based localization techniques for users in low and high mobility scenarios of a well mapped ray tracing DeepMIMO dataset 
\item Hybridization of adaptive filtering technique using EKF and UKF filter to dynamically switch between scenarios based on user's speed.

\end{enumerate}

The remainder of this paper is structured as follows: Section II presents the system model and proposed adaptive hybrid filtering framework. Section III details the methodology and the implementation of adaptive filtering techniques for the hybrid localization model. In section IV, the results and evaluation metrics are discussed, as well as a comparison with the state-of-the-art model. Finally, Section V summarizes the key findings and outlines possibilities for future research.

\section{System Model}

We model two outdoor scenarios using a single mmWave channel as shown in Fig. 1. In the channel, a user equipment moving at a pedestrian speed and a vehicle moving at high speed are under the influence of signals from the transmitting base stations (gNB) with gNB1 as physical anchor, gNB2 and gNB3 serving as virtual anchors. Reflectors such as walls and trees provide non-line of sight (NLoS) multipath for the users, but the NLoS paths are not considered in this work. The gNB carrier frequency, $f_{\mathrm{c}} = 28\,\mathrm{GHz}$ with signal bandwidth $B = 100\,\mathrm{MHz}$; the channel therefore supports centimetre‑resolution range and sub‑degree angular estimation. The sensing modalities of the transmitting signal are adopted in the development of the hybrid localization technique for both the pedestrian user (low mobility user) and the high mobility vehicles. And the Bayesian filters (particularly the EKF and UKF) adopted in the modeling and estimation of user trajectories in the two speed mode have the following state propagation, prediction and measurement models.


\subsection{Kinematic State Definition and Motion Model}\label{sec:unified_state}
Discrete epochs are indexed by $k \in \{0,1,\dots,N\}$ with sampling interval $\Delta t = 1\,\mathrm{s}$, as given in the DeepMIMO metadata adopted in this work.

The state propagation vector for both low-mobility (LM) and high-mobility (HM) users is defined as a 3D kinematic model that captures the user's position, velocity, and oscillator bias at each time step $k$. The user position vector at time $k$ is given by $\mathbf{p}_k = [x_k,\, y_k,\, z_k]^{\top}$ in meters $(\mathrm{m})$, and the corresponding velocity vector is $\mathbf{v}_k = [v_{x,k},\, v_{y,k},\, v_{z,k}]^{\top}$ in meters per second $(\mathrm{m\,s}^{-1})$. A slowly varying oscillator or Doppler bias is denoted by $b_k$ in hertz $(\mathrm{Hz})$.

The full state vector $\mathbf{x}_k$ for both LM and HM scenarios is therefore defined as:
\begin{equation}
\mathbf{x}_k = 
[x_k;\, y_k;\, z_k;\, v_{x,k};\, v_{y,k};\, v_{z,k};\, b_k]^{\top}
\in \mathbb{R}^{7}
\end{equation}

This unified 3D formulation allows consistent modeling across varying mobility conditions. For flat terrain (e.g., pedestrians on the ground), $z_k$ and $v_{z,k}$ can be set to zero; for vehicular or aerial mobility, vertical dynamics are captured explicitly.

\paragraph*{State Propagation Model}
The user follows a constant-velocity (CV) motion model discretized using a zero-order hold. The state transition equation is:
\begin{equation}
\mathbf{x}_{k+1} = F \mathbf{x}_k + \mathbf{w}_k,
\quad \mathbf{w}_k \sim \mathcal{N}(\mathbf{0}, Q_k)
\label{eq:cv_model}
\end{equation}
where $F$ is the state transition matrix and $Q_k$ is the process noise covariance matrix, adapted per epoch.

\paragraph*{State Transition Matrix $F$}
\begin{equation}
F = 
\begin{bmatrix}
I_3 & \Delta t\, I_3 & \mathbf{0}_{3\times 1} \\\\
\mathbf{0}_{3\times 3} & I_3 & \mathbf{0}_{3\times 1} \\\\
\mathbf{0}_{1\times 3} & \mathbf{0}_{1\times 3} & 1
\end{bmatrix}
\end{equation}
where $I_3$ is the 3 by 3 identity matrix, and $\Delta t$ is the sampling interval (typically $1\,\mathrm{s}$). This structure models constant velocity in 3D space with independent evolution of the bias term $b_k$.

\paragraph*{Process Noise Covariance $Q_k$}
To capture uncertainties in user motion and oscillator drift, we define a structured process noise covariance matrix. The position and velocity variances account for lateral and vertical mobility, while the oscillator bias term reflects frequency instability. The Doppler spread $\sigma_{\mathrm{D},k}$ at each epoch $k$ governs the velocity variance through proportional scaling.
\begin{equation}
Q_k = \mathrm{diag}(\sigma_p^2,\, \sigma_p^2,\, \sigma_{p,z}^2,\,
                   \sigma_v^2,\, \sigma_v^2,\, \sigma_{v,z}^2,\, \sigma_b^2),
\quad \sigma_v^2 \propto \sigma_{\mathrm{D},k}
\end{equation}
\textbf{Note:} The horizontal motion noise is assumed symmetric: $\sigma_{p,x}^2 = \sigma_{p,y}^2 = \sigma_p^2, \quad \sigma_{v,x}^2 = \sigma_{v,y}^2 = \sigma_v^2$ while vertical noise may differ: $\sigma_{p,z}^2 \ne \sigma_p^2, \quad \sigma_{v,z}^2 \ne \sigma_v^2.$
For low-mobility users, vertical motion components $(z_k,\, v_{z,k})$ can be fixed to zero without affecting filter consistency.

\subsection{Measurement Equations: ToA, AoA, AoD, and Doppler}
\label{sec:meas}

At each sensing epoch $k$, the user equipment (UE) receives joint communication-sensing signals from a set of calibrated base stations (gNBs), comprising one line-of-sight (LoS) anchor ($\ell=1$) and multiple virtual reflectors ($\ell=2,3$). Each anchor $\ell$ has a known location $\mathbf{p}^{(\ell)} = [x^{(\ell)},\, y^{(\ell)},\, z^{(\ell)}]^\top \in \mathbb{R}^3$. The user's position is denoted by $\mathbf{p}_k = [x_k,\, y_k,\, z_k]^\top$ and velocity by $\mathbf{v}_k = [v_{x,k},\, v_{y,k},\, v_{z,k}]^\top$ at time step $k$.

The gNB extracts the following measurements:
{Time-of-Arrival (ToA)} $\tau_k^{(\ell)}$;{Angle-of-Arrival (AoA)} $\alpha_k^{(\ell)}$ and {Angle-of-Departure (AoD)} $\theta_k^{(\ell)}$; {Line-of-Sight Doppler shift} $d_k$.
For low-mobility UEs (e.g. pedestrians, quasi-static motion), additional odometry readings such as translational speed and heading $(v_{\text{odo},k},\, \phi_{\text{odo},k})$ may be available.

The nonlinear measurement models used in this work relate the physical quantities sensed by the gNB to the user’s spatial state. Each observation channel captures a different aspect of the environment–user interaction:

First, the \emph{ToA} from anchor $\ell$ provides a range-based constraint and is modeled as:
\begin{equation}
\tau_k^{(\ell)} 
  = \frac{\lVert \mathbf{p}_k - \mathbf{p}^{(\ell)} \rVert}{c} + w_{\tau},
\label{eq:toa}
\end{equation}
where $\lVert \mathbf{p}_k - \mathbf{p}^{(\ell)} \rVert$ is the Euclidean distance between the user and anchor, $c$ is the speed of light, and $w_{\tau}$ is Gaussian timing noise.

Second, the \emph{AoA} measured from anchor~$\ell$
provides the UE’s azimuthal bearing in the horizontal ($x$–$y$) plane of the three-dimensional coordinate frame:
%
\begin{subequations}\label{eq:aoa_3d}
\begin{align}
\alpha_k^{(\ell)} &=
  \operatorname{atan2}\!\bigl(y_k-y^{(\ell)},\,x_k-x^{(\ell)}\bigr) + w_{\alpha},\label{eq:azimuth}\\[2pt]
\beta_k^{(\ell)} &=
  \operatorname{atan2}\!\bigl(z_k-z^{(\ell)},\,
           \sqrt{(x_k-x^{(\ell)})^{2}+(y_k-y^{(\ell)})^{2}}\bigr)
  + w_{\beta},\label{eq:elevation}
\end{align}
\end{subequations}

Here $\alpha_k^{(\ell)}$ is the \emph{azimuth} (bearing in the horizontal
$x$–$y$ plane) and $\beta_k^{(\ell)}$ is the \emph{elevation} above that
plane; $(x_k,y_k,z_k)$ and $(x^{(\ell)},y^{(\ell)},z^{(\ell)})$ denote the UE
and anchor coordinates, respectively, while $w_{\alpha}$ and $w_{\beta}$ are
zero-mean Gaussian angle-measurement noises.

Third, the \emph{AoD} captures the direction from the transmitter to the user and is modeled similarly as:
\begin{equation}
\theta_k^{(\ell)} 
  = \operatorname{atan2}(y^{(\ell)} - y_k,\, x^{(\ell)} - x_k) + w_{\theta},
\end{equation}
where $w_{\theta}$ is the associated departure angle noise. AoD is particularly useful for cooperative gNB-side localization.

Lastly, the \emph{Line-of-Sight (LoS) Doppler shift} $d_k$ encodes the radial velocity between the user and LoS anchor $\ell=1$:
\begin{equation}
d_k 
  = \frac{\mathbf{v}_k^\top(\mathbf{p}_k - \mathbf{p}^{(1)})}
           {\lVert \mathbf{p}_k - \mathbf{p}^{(1)} \rVert}
     + b_k + w_d,
\label{eq:doppler}
\end{equation}
where $\mathbf{v}_k$ is the user's velocity vector, $b_k$ is the Doppler bias due to oscillator drift, and $w_d$ models frequency estimation uncertainty.
These equations jointly encode geometric range, bearing angles, and Doppler shifts, forming the observation model that feeds the Kalman filter. The fused measurement vector dynamically adapts to UE mobility, allowing robust filtering and trajectory estimation in the proposed hybrid localization framework.

\begin{figure}[!t]            
  \centering
  \includegraphics[width=\linewidth]{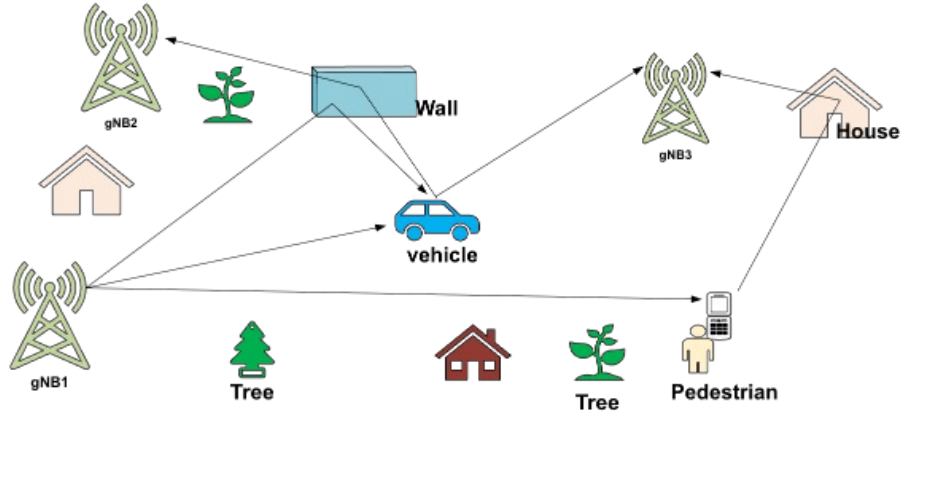}
  \caption{Illustration of a mixed-mobility bistatic mmWave localization scene. gNB1 represent the physical anchor, gNB2 and gNB3 are virtual anchors. The wall, trees, and buildings serve as reflectors. A pedestrian (low mobility) and a car (high mobility) are localized through direct and reflected paths.}
  \label{fig:sysmodel}
  \vspace{-0.4em}
\end{figure}

\begin{figure}[!t]
  \centering
  \includegraphics[width=0.95\linewidth]{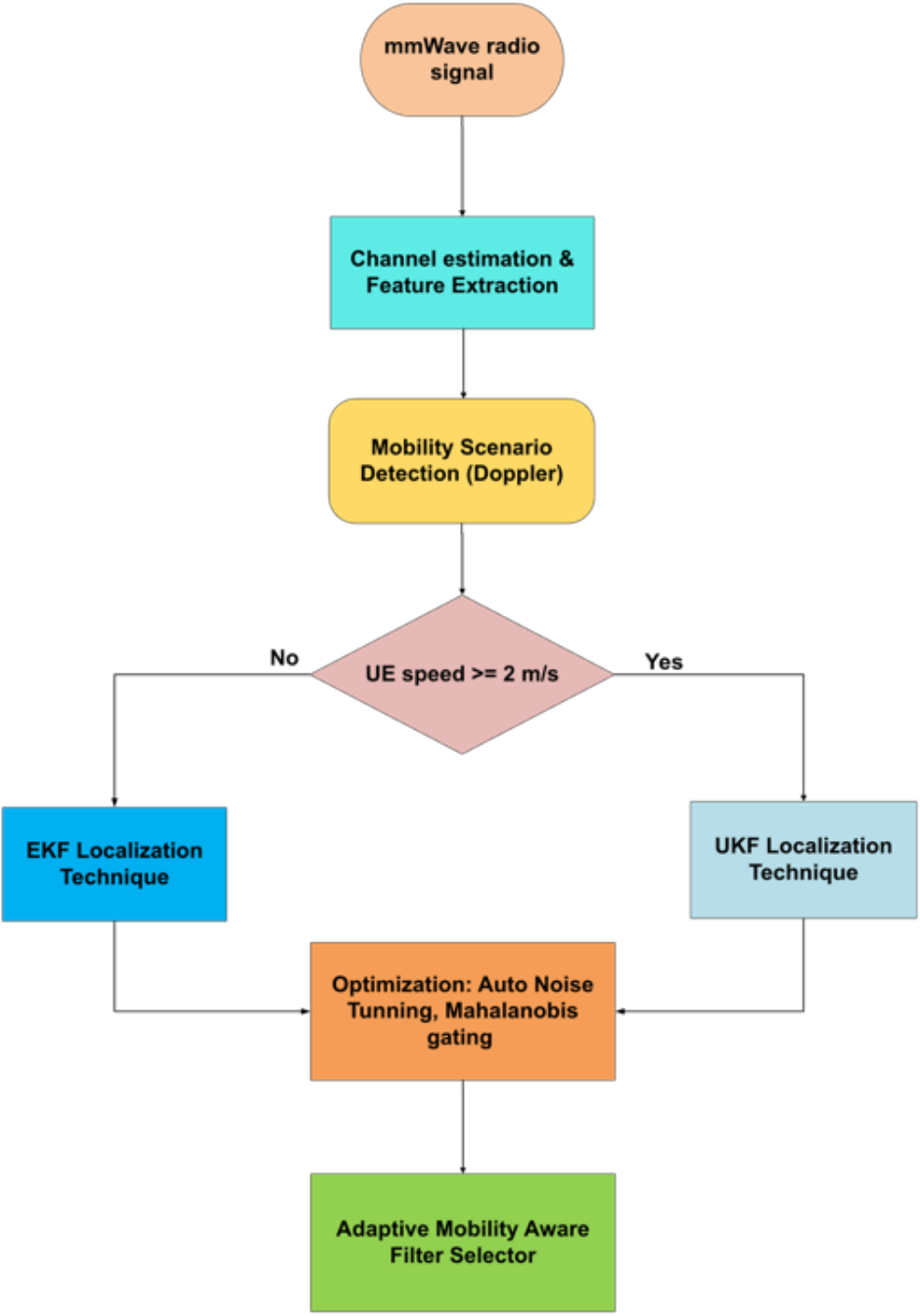}
  \caption{Mobility-aware Adaptive localization model.}
  \label{fig:hybrid}
  \vspace{-0.4em}
\end{figure}

\section{Methods: Adaptive Filtering Approach}

This section outlines the design implementation of the hybrid localization framework and mobility-aware trajectory estimation using the adaptive hybrid filtering. The experimental design leverages the ray-traced DeepMIMO dataset\cite{alkhateeb2019deepmimo}, integrating both low mobility and high mobility environments to reflect real-world deployment scenarios.As illustrated in Fig. 2, we simulate two experimental conditions within a single environment: the active condition is determined by the user’s speed. Two complementary filters are used to localize users in these scenarios. The scenarios are drawn from the DeepMIMO dataset and simulated in MATLAB; they are as follows:
\begin{itemize}
  \item \emph{Scenario~1 (outdoor 1)} represents a pedestrian-level, low mobility setting, emphasizing quasi-static dynamics.
  \item \emph{Dynamic Doppler Scenario} emulates high-speed vehicular motion, emphasizing nonlinear kinematics and Doppler spreads.
\end{itemize}
The AoA, AoD, and ToA parameters extracted from the mmWave channel, along with the Doppler shift, are used for localization in low mobility and high mobility scenarios. The Bayesian filters; EKF, UKF and Cubarture Kalman Filter (CKF) are tested on each scenario datasets to estimate the UE trajectory and their performance are optimized for the localization technique. The three filters tested responded differently to the changes in the channel and UE dynamics (see Tables I and II). The best performing filters (EKF and UKF) from each scenario is selected and adapted for the mobility aware localization, and their fusion defines the adaptive hybrid filtering model. These two filters (EKF and UKF) handles the estimation error well in their respective scenario and they are implemented as adaptive filtering in the mobility aware localization technique. 
The optimization techniques such as Mahalanobis gating, adaptive noise scaling, and Rauch–Tung–Striebel (RTS) smoothing are applied to maximize filter performance highlighted in Sec. IV. 
\subsection{Mobility Classifier and Filter Switching}
The UE speed determines the switch between the localization technique as shown in the Fig. 2.
At each time step~$k$, the system estimates the user’s speed as:
\begin{equation}
  \hat v_k = \bigl\lVert \hat{\mathbf v}_{k|k-1} \bigr\rVert.
\end{equation}
When $\hat{v}_k \le  2\,\mathrm{m/s}$ based on the UE speed model in the adopted dataset, the localization technique applies a low-mobility EKF technique; otherwise, the high-mobility UKF technique is activated. This adaptive switching ensures optimal filter selection under changing kinematics.
\subsection{Prediction Step: 3-D Constant-Velocity Propagation}
At each epoch, the filter advances the full 3-D kinematic state under a discrete
constant–velocity (CV) assumption.  Denote the predicted speed vector by
$\hat{\mathbf v}_{k|k-1}=[\hat v_{x,k}\;\hat v_{y,k}\;\hat v_{z,k}]^{\!\top}$
and the state dimension by $n$ (seven for EKF, identical for the UKF branch).

\begin{align}
  \hat{\mathbf x}_{k|k-1} &= F\,\hat{\mathbf x}_{k-1|k-1},\\
  P_{k|k-1}              &= F\,P_{k-1|k-1}\,F^{\top}+Q_k,
\end{align}
where the 3-D CV transition matrix is
\begin{equation}
  F=
  \begin{bmatrix}
    I_3 & \Delta t\,I_3 & \mathbf 0_{3\times1}\\[2pt]
    \mathbf 0_{3\times3} & I_3 & \mathbf 0_{3\times1}\\[2pt]
    \mathbf 0_{1\times3} & \mathbf 0_{1\times3} & 1
  \end{bmatrix}\!,
\end{equation}
and $Q_k=\operatorname{diag}(\sigma_{p}^2,\sigma_{p}^2,\sigma_{p,z}^2,\sigma_{v}^2,\sigma_{v}^2,\sigma_{v,z}^2, \sigma_{b}^2)$.
To keep the process noise consistent with channel conditions, the velocity variance is scaled by the epoch-specific Doppler spread, $\sigma_{v}^{2}$.
For the UKF branch, the \emph{unscented transform} builds the sigma set
$\{\chi_{k-1}^{(i)}\}_{i=0}^{2n}$, advances each point with
$\chi_{k|k-1}^{(i)} = F\,\chi_{k-1}^{(i)}$, and fuses the results into the
predicted mean and covariance.  
This second-order prediction respects the Doppler-adaptive $Q_k$ and remains
accurate across both pedestrian and vehicular mobility regimes.
\subsection{Measurement Update}

\subsubsection{EKF branch}  
The nonlinear observation model in  \eqref{eq:toa}--\eqref{eq:doppler} is
first linearized about the prediction $\hat{\mathbf x}_{k|k-1}$ to obtain the
Jacobian $H_k$.  With the innovation
$\mathbf y_k=\mathbf z_k-\hat{\mathbf z}_{k|k-1}$, the Kalman gain and posterior estimates follow:
\begin{IEEEeqnarray}{rCl}
K_k      &=& P_{k|k-1}H_k^{\top}\!\bigl(H_k P_{k|k-1} H_k^{\top}+R_k\bigr)^{-1},\\
\hat{\mathbf x}_{k|k} &=& \hat{\mathbf x}_{k|k-1}+K_k\,\mathbf y_k,\\
P_{k|k}               &=& \bigl(I-K_k H_k\bigr)P_{k|k-1}.
\end{IEEEeqnarray}

\subsubsection{UKF branch}  
Each sigma point $\chi_{k|k-1}^{(i)}$ is passed through the nonlinear measurement function to obtain $\zeta_{k}^{(i)}=h\!\bigl(\chi_{k|k-1}^{(i)}\bigr)$.  The weighted set $\{\zeta_{k}^{(i)}\}$ yields the predicted measurement mean $\hat{\mathbf z}_{k|k-1}$, innovation covariance $S_k$, and gain 
\begin{equation}
    K_k = P_{xz,k}\,S_k^{-1},
\end{equation}
where $P_{xz,k}$ is the cross-covariance between state and measurement sigma
points.  The posterior state and covariance are then updated exactly as in the
EKF expressions above, but with the unscented $K_k$.

\subsection{Outlier Rejection via $\chi^2$ Gating}
To screen spurious observations before the correction step, the algorithm computes the \emph{normalized innovation squared} (NIS) i.e., the squared Mahalanobis distance of the innovation vector:
$d_k^{2} = \mathbf y_k^{\top} S_k^{-1} \mathbf y_k,\;
\mathbf y_k = \mathbf z_k - \hat{\mathbf z}_{k|k-1}$,
where $\mathbf y_k$ is the measurement residual and
$S_k=H_kP_{k|k-1}H_k^{\top}+R_k$ (or its unscented equivalent) is the
innovation covariance.  A measurement is accepted only if
$d_k^{2}$ remains below the 99\,\% percentile of the
$\chi^{2}$ distribution with $d=\dim(\mathbf y_k)$ degrees of freedom; any observation violating this threshold is treated as an outlier and discarded.
\subsection{Adaptive Noise Scaling}
The filter tracks the \emph{normalized innovation squared} $\text{NIS}_k=d_k^2/d$, where $d_k^2=\mathbf y_k^{\top}S_k^{-1}\mathbf y_k$
and $d=\dim(\mathbf y_k)$.  When the window‐averaged
$\mathbb{E}[d_k^2]$ drifts from its theoretical mean~$d$, the process-noise covariance $Q_k$ and measurement-noise covariance $R_k$ are
jointly rescaled by a factor $\gamma\!\in\![0.5,2]$, keeping the estimator statistically consistent under changing SNR and Doppler spread.

\subsection{Rauch--Tung--Striebel Smoothing}
After every $N_w$ epochs (e.g.\ $N_w=200$), a backward RTS pass refines the entire state trajectory:
\begin{align}
C_k &= P_{k|k}F^{\top}\!\bigl(FP_{k|k}F^{\top}+Q_k\bigr)^{-1},\\
\hat{\mathbf x}_{k|N} &= \hat{\mathbf x}_{k|k}+C_k
\bigl(\hat{\mathbf x}_{k+1|N}-\hat{\mathbf x}_{k+1|k}\bigr),
\end{align}
where $C_k$ is the \emph{smoothing gain} and $\hat{\mathbf x}_{k|N}$ the smoothed state at epoch~$k$ (\(N\) = final time). Exploiting future measurements in this way typically halves the ATE relative to the forward filter alone.
\subsection{Adaptive Mobility-Aware Localization Model} The proposed \emph{hybrid localization scheme} (outlined in Algorithm 1) integrates both EKF and UKF techniques within a single, velocity–adaptive localization pipeline. The two Bayesian filters (EKF and UKF) were selected based on the performance metrics specified in the next section.
A \textit{speed-gate} monitors the odometry-derived speed $\lVert \hat v_k\rVert$ and  

\begin{itemize}
  \item activates the \textbf{EKF} when
$\lVert\hat v_k \rVert\le 2\,\mathrm{m/s}$, exploiting the near-linearity of pedestrian motion for maximal computational efficiency;
  \item switches to the \textbf{UKF} for $2 < \lVert \hat v_k\rVert \le 20\,\mathrm{m/s}$, maintaining estimator consistency under the stronger non-linear dynamics of vehicular travel.
\end{itemize}

Both branches share three key mechanisms:   
(i) innovation-based adaptive $Q/R$ scaling,  
(ii) $\chi^{2}$ Mahalanobis gating to reject multipath outliers, and  
(iii) an RTS smoother that back-propagates information to produce a globally consistent trajectory.

\begin{algorithm}
\caption{Mobility-Aware Hybrid Localization Technique}
\LinesNotNumbered        
\SetAlgoNoLine           
\KwIn{AoA, AoD, ToA, Doppler shift}
\KwOut{Localization Technique: EKF or UKF Model}

\textbf{Step 1: Acquire speed}\\
\If{$v_{\text{user}}$ \textbf{is not provided}}{
  prompt ``Enter current speed (m/s):''\;
  \lIf{input \emph{not numeric}}{\textbf{show error}}
  $v_{\text{user}} \gets$ input value\;
}

\smallskip
\textbf{Step 2: Define thresholds}\\
$v_{\mathrm{LM}} \gets 2.0$ m/s,\;
$v_{\mathrm{HM}} \gets 20.0$ m/s\;

\smallskip
\textbf{Step 3: Filter selection}\\
\lIf{$v_{\text{user}} \le v_{\mathrm{LM}}$}{
  \textbf{run} \emph{EKF\_Pipeline()} $\rightarrow$ metrics
}
\lElseIf{$v_{\mathrm{LM}} < v_{\text{user}} \le v_{\mathrm{HM}}$}{
  \textbf{run} \emph{UKF\_Pipeline()} $\rightarrow$ metrics
}
\lElse{
  \textbf{warn}: speed outside design window; abort or fallback
}

\smallskip
\textbf{Step 4: Return results}\\
\Return{Performance Evaluation Result (ATE,RPE,NEES,RMSE)}

\end{algorithm}

\section{Result Evaluation and Discussion}
\noindent
The evaluation results of the developed localization models are highlighted in Tables I and II, and the performance metrics used in the evaluation are as follows: \\
\textit{Absolute Trajectory Error}: It captures the drift in the estimated trajectory with respect to a ground-truth path and is therefore a primary indicator of long-term navigation accuracy. The State-of-the-art work on indoor and low-mobility localization models mostly reported \emph{sub-metre} ATE between 0.3m and 1m, whereas ATE for vehicles and UE in high-mobility scenarios ranges between 1 to 2m depending on speed, multipaths, and scatterers~\cite{shahmansoori2017position, jamaludin2025slam}.
\\
\textit{Root Mean Square Error of ATE}: This metric quantifies how reliably the filter keeps the estimated path aligned with the true path. A smaller RMSE-ATE therefore indicates a steadier, more consistently accurate trajectory solution.\\
\textit{Relative Pose Error}: RPE measures short-horizon drift between successive frames (e.g., over \(1\;\text{s}\)) and reflects how smoothly the filter tracks motion\cite{yi2019metrics}.
Existing work shows per-second translation drift of only \(0.01\text{–}0.05\)m for low mobility; pedestrian modeled datasets and about \(0.1\text{–}0.3\)m for vehicular and high mobility speeds~\cite{jamaludin2025slam}.
\\
\textit{Normalized Estimated Error Squared}: This quantifies filter consistency while indicating the alignment between the true error and the covariance predicted by the estimator.
For an \(n\)-dimensional state, a well-tuned (optimized) filter satisfies \(\mathbb{E}[\text{NEES}] \approx n\); for the \(3\)-D position it implies a mean of \(\text{NEES} \approx 3\) under a \(\chi^{2}_{2}\) distribution, and a \(95\,\%\) gate at \(5.99\). 
Values far \emph{above} this bound indicate over-confidence (risk of divergence), while values far \emph{below} suggest an overly pessimistic covariance.

\begin{table}[t]
  \caption{Evaluation Results of Filter Performance (Low-Mobility Scenario)}
  \label{tab:tuned_filter_results}
  \centering
  \renewcommand{\arraystretch}{1.1}  
  \begin{tabular}{lcccc}
    \hline                              
    \textbf{Filter} &
    \textbf{ATE [m]} &
    \textbf{RPE [m]} &
    \textbf{NEES} &
    \textbf{RMSE [m]} \\ \hline         
    EKF & 0.67 & 0.10 & 2.40$\times$10\textsuperscript{-4}    & 0.30 \\
    UKF & 0.35 & 0.11 & 6779.6   & 0.22 \\
    CKF & 2.72 & 0.11 & 2.06 $\times$10\textsuperscript{5}  & 1.24 \\ 
   Optimized EKF & 0.19 & 0.02 & 0.75     & 0.06\\
    \hline   
  \end{tabular}
\end{table}

\begin{table}[t]
  \caption{Evaluation Results of Filter Performance (High-Mobility Scenario)}
  \label{tab:filter_results}
  \centering
  \renewcommand{\arraystretch}{1.1}  
  \begin{tabular}{lcccc}
    \hline
    \textbf{Filter} &
    \textbf{ATE [m]} &
    \textbf{RPE [m]} &
    \textbf{NEES} &
    \textbf{RMSE [m]} \\ \hline
    EKF & 158.3   & 2.751  & 1 704.3    & 198.76 \\
    UKF & 8.69    & 33.99   & 5.08$\times$10\textsuperscript{5} & 22.32 \\
    CKF & 12.868  & 4.684  & 2.59$\times$10\textsuperscript{7} & 47.72 \\ 
    Optimized UKF & 1.74    & 0.17   & 1.36$\times$10\textsuperscript{5} & 0.60 \\
    \hline
  \end{tabular}
\end{table}

Tables I and II show evaluation outcomes of the developed localization techniques for low and high mobility scenarios, respectively. 
The three filters (EKF, UKF, and CKF) adopted in developing the technique for each scenario and their performance as evaluated by the earlier stated metrics are shown in these tables. From the tables, the \text{optimized EKF} and \text{UKF} markedly outperform the cubature Kalman filter (CKF).
For \emph{low-mobility} trajectories, the EKF attains an \textbf{ATE $=0.19\,$}m, whereas the UKF maintains high accuracy under \emph{high-mobility} conditions with \textbf{ATE $=1.74\,$}m.  
Both outcomes surpass the reported state-of-the-art results by sizable margins, confirming the benefit of the proposed hybrid design.

\subsection{Model comparison Against the State of the Art}
In the work by \cite{karttunen2025towards}, the author developed a technique for localizing the user’s device within an indoor environment using a bistatic radio signal in the mmWave channel. The developed model has the ATE between 0.2m to 0.3m. The performance metrics of three localization algorithms for two mobile robots were evaluated in \cite{scheideman2020flexible}. The ATE of the localization algorithm for the two robots is between 0.016 m and 0.317 m.
New benchmark metrics tagged distance error and loop closure detection were used for assessing the performance of LIDAR-localization techniques within an outdoor environment, as highlighted in \cite{jamaludin2025slam}. The values of these metrics are between 0.11m and 0.71m. The author deployed the use of four popular algorithms (RTAB-Map, LIO-SAM, FAST-LIO and NeuroSLAM) within an oil plantation environment during different times of the day. 
RadarSLAM in~\cite{hong2020radarslam} uses frequency-modulated radar for localization and mapping in adverse weather. The technique reported an ATE in the 5-13m range over long routes, with RPE $\sim 2\%$ translation drift. Table~III compares our developed model with these state-of-the-art outcomes. The developed techniques exhibit significantly lower ATE and RPE over shorter paths, with consistent NEES, for both the low mobility model and the high mobility model, compared to the metric results of the indoor environment and the outdoor environment model.
\begin{table}[!t]                  
  \caption{Comparison with mmWave/SLAM localization models}
  \label{tab:ate_comparison}
  \centering
  \footnotesize                     
  \setlength{\tabcolsep}{4pt}       
  \renewcommand{\arraystretch}{1.1}
  \begin{tabular}{llcc}
    \hline
    \textbf{Method} & \textbf{Scenario} & \textbf{ATE [m]} & \textbf{Improvement}\\ \hline
    Indoor mmWave\,\cite{karttunen2025towards} & Indoor                & 0.2--0.3 & Baseline\\
    RadarSLAM\,\cite{hong2020radarslam}     & Outdoor (adverse)     & 5--13    & Lower perf.\\
    Proposed EKF              & Low-mob.\ outdoor     & \textbf{0.19} & Significant\\
    Proposed UKF              & High-mob.\ outdoor    & \textbf{1.74} & $\sim$60\% better\\ \hline
  \end{tabular}
\end{table}

\FloatBarrier
\section{Conclusion}

We introduced a \emph{mobility-aware hybrid localization} technique that dynamically switches between optimized EKF and UKF filters according to the user’s velocity in mmWave channels.  
This adaptive model substantially enhances accuracy, sustaining an ATE below~$<\!0.25\,\text{m}$ for pedestrian motion and about~$2\,\text{m}$ for vehicular speeds. The optimization techniques, such as adaptive covariance scaling, RTS smoothing, and $\chi^2$ innovation gating, ensure robustness against outliers and model mismatch.  
Overall, the hybrid approach delivers \mbox{30–60\%} reduction in localization error compared with state-of-the-art single filter techniques, offering a
practical path toward next-generation beam management solutions using localization.
In our future work, localization will be carried out in an unmapped environment using the SLAM technique.  We will also evaluate other types of filters, (such as PHD, RBPF, etc.,) using multimodal sensing system.

\section*{Acknowledgment}
This work was jointly supported by the African Center of Excellence in Internet of Things (ACEIoT), College of Science and Technology, University of Rwanda, Partnership for Skills in  Applied Sciences, Engineering \& Technology (PASET), Regional Scholarship \& Innovation Fund (RSIF), and National Research Foundation of Korea (NRF) grant funded by the Korea government (MSIT) (No. RS-2024-00409492). 

\bibliographystyle{IEEEtran}
\bibliography{IEEEabrv, Reference}
\end{document}